\documentclass[graybox]{svmult}

\usepackage{type1cm}        
\usepackage{makeidx}         
\usepackage{graphicx}        
\usepackage{multicol}        
\usepackage[bottom]{footmisc}

\usepackage{hyperref}
\usepackage{etoolbox}
\appto\UrlBreaks{\do\-}
\usepackage{newtxtext}       %
\usepackage{newtxmath}       

\makeindex             

\begin{document}

\title*{DroidDissector: A Static and Dynamic Analysis Tool for Android Malware Detection}

\author{Ali Muzaffar, Hani Ragab Hassen, Hind Zantout and Michael A Lones}
\institute{Ali Muzaffar \at Heriot-Watt University, Dubai, UAE, \email{am29@hw.ac.uk} \and Hani Ragab Hassen \at Heriot-Watt University, Dubai, UAE, \email{h.ragabhassen@hw.ac.uk} \and Hind Zantout \at Heriot-Watt University, Dubai, UAE, \email{h.zantout@hw.ac.uk} \and Michael A Lones \at Heriot-Watt University, Edinburgh EH14 4AS, United Kingdom, \email{m.lones@hw.ac.uk}  }
%
%
\maketitle

\abstract*{DroidDissector is an extraction tool for both static and dynamic features. The aim is to provide Android malware researchers and analysts with an integrated tool that can extract all of the most widely used features in Android malware detection from one location. The static analysis module extracts features from both the manifest file and the source code of the application to obtain a broad array of features that include permissions, API call graphs and opcodes. The dynamic analysis module runs on the latest version of Android and analyses the complete behaviour of an application by tracking the system calls used, network traffic generated, API calls used and log files produced by the application.}

\abstract{DroidDissector is an extraction tool for both static and dynamic features. The aim is to provide Android malware researchers and analysts with an integrated tool that can extract all of the most widely used features in Android malware detection from one location. The static analysis module extracts features from both the manifest file and the source code of the application to obtain a broad array of features that include permissions, API call graphs and opcodes. The dynamic analysis module runs on the latest version of Android and analyses the complete behaviour of an application by tracking the system calls used, network traffic generated, API calls used and log files produced by the application.}

\section{Introduction}
\label{sec:1}

Android has continued to evolve and gain popularity. According to statscounter \cite{androidData}, it holds a share of 71.74 \% of the smartphone market, 44.11 \% more than its major competitor iOS in 2023. One of the major reasons behind its popularity is the ease of access to Android applications, whether using the official Google Play Store or third party application stores. 

Android's popularity brings with it a number of security concerns, malware being a major one. Malware is any software or application developed with the intention to harm the device it is installed on. The consequences of this can include monetary loss, information leakage and advertising spam. According to G Data Mobile Security Report \cite{malwareStat} released in 2022, a new piece of Android malware is observed every 23 seconds. Therefore, it is essential to have a mechanism to detect malware before it causes any harm. 

Traditionally, anti-malware techniques have used signature based malware detection. However, this mechanism requires maintaining a known malware database, and detecting new malware is a challenge. Recent years have seen an increase in the popularity of machine learning based approaches to malware detection, and these have shown a lot of potential \cite{Muzaffar2022}. Generally, these approaches require the collection of a dataset of Android applications, followed by the extraction of features using static or dynamic analysis. Our dynamic and static analysis tool (DroidDissector) extracts the most used and most useful static and dynamic features, as identified in our recent survey paper \cite{Muzaffar2022}.

The most commonly used dynamic analysis tool in the literature is DroidBox \cite{droidbox} which works on Android 4.1.2, released a decade ago. The tool extracts information like SMS sent and received, read and write operations, cryptographic operations, and information leakage. The author does not maintain the tool anymore and the latest GitHub commit is over 3 years ago.  AndroPyTool \cite{andropy1,andropy2} extracts both static and dynamic features and stores the information in JSON files, CSV files, or writes the output to a database. The dynamic analysis module uses DroidBox and therefore only works on versions of Android up to 4.1.2. AndroPyTool also extracts system call traces in addition to the features extracted by DroidBox. Currently the most up to date and supported analysis tool is Mobile Security Framework (MobSF). The dynamic analysis module can run on both a third party virtual environment, Genymotion, or the Android virtual device which comes with Android SDK \cite{mobsf}. MobSF is aimed at professional environments and is not commonly used in research.  

In this paper, we present DroidDissector \footnote{\url{https://github.com/alisolehria/DroidDissector-A-Static-and-Dynamic-Analysis-Tool-for-Android-Malware-Detection}}, a fully integrated static and dynamic analysis tool for extracting features; these can then be used for Android malware detection or analysing the behaviour of an application. We developed this tool to address the limitations of currently available static and dynamic analysis tools, notably the fact that existing static analysis tools generally only extract a single type of feature (such as permissions or API calls) and that dynamic analysis tools are in most cases outdated or focus on specific commercial use cases. DroidDissector allows users to run an analysis with one central tool and can extract multiple features. The tool runs on the most up to date Android version.  We used DroidDissector in two of our previous works.  In \cite{Muzaffar2021}, we carried out an investigation on API calls and how feature selection and machine learning models affect detection rates. The results showed an accuracy rate of up to 96\%. We also carried out an extensive study on features for Android malware detection in \cite{Muzaffar2023} and used DroidDissector to extract static and dynamic features.

\section{Static Analysis Tool}

\begin{figure}[t]
\includegraphics[scale=0.5]{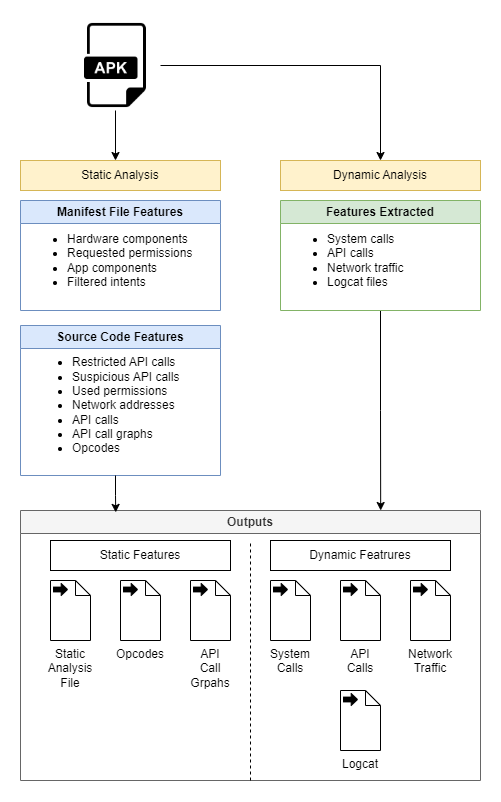}
\centering
\caption{DroidDissector Architecture}
\label{fig:tool}
\end{figure}

Static features are the most widely used features in Android malware detection.
The static analysis module of DroidDissector reverse engineers and extracts the manifest file and smali files from an APK package using ``APKtool" \cite{APKTool}. APKTool creates a directory for each APK file which contains the reverse engineered smali and manifest files. For each application, DroidDissector then extracts features and stores then in three separate files:

\begin{itemize}
    \item Static Features file: This is the main file produced by the static analysis module. This file includes features from both the manifest and smali files. The features saved in this file include:
    \begin{itemize}
        \item Hardware Components - An application can use several hardware components of the device such as camera and GPS and are defined in the manifest file. 
        \item Requested Permissions - Android security revolves around permissions. Permissions are part of the Android sandboxing approach and allow different applications to interact with each other and the system. Permissions used by an application are defined in the manifest file. 
        \item App Components - An Android application consists of four components: activities, services, broadcast receivers, and content providers and are defined in the manifest file. 
        \item Filtered Intents - Filtered intents allow applications to specify the intents it would like to receive, also defined in the manifest file. 
        \item Restricted API Calls - Drebin \cite{Arp2014} shortlisted restricted API calls for malware detection extracted from the smali files produced by APKTool. 
        \item Suspicious API Calls - Drebin \cite{Arp2014} shortlisted suspicious API calls for malware detection. DroidDissector analyzes the smali files extracted from the APK to extract this feature.
        \item Used Permissions - A number of permissions can be listed in the manifest file. However, this does not mean the application will use them. The static analysis module extracts the permissions that are actually used by analysing the smali files. 
        \item Network Addresses - Network addresses present in the smali code. 
        \item API Calls v30 - The static analysis module of DroidDissector also mines for any API calls that are used by the application by analysing the smali code. 
    \end{itemize}
    \item Opcodes file: Opcodes are the low-level operations executed by the Android virtual machine. DroidDissector then stored the sequence of opcodes used by the application in an opcode file.
    \item API Call Graphs: We used FLOWDROID \cite{Arzt2014} to extract call graphs produced by the applications. DroidDissector stores the API call graphs in API call graphs files which can later be used to obtain the sequence of API calls which are called by the application. 
\end{itemize}

The primary language used to build the static analysis module was Python 3.8.  DroidDissector performs static analysis of each APK file in Python, except for the extraction of API call graphs, which uses FLOWDROID, a Java library.

\section{Dynamic Analysis Tool}

We tested the dynamic analysis tool on Linux distributions including Ubuntu and CentOS, as well as Windows operating system. DroidDissector’s dynamic analysis tool requires the following to run:

\begin{itemize}
    \item Python 3: The tool requires python 3 to run. 
    \item Frida-Android: Frida's python library is required to hook to API calls during execution of an application.
    \item Android Emulator: The dynamic analysis module uses the virtual environment provided by Android SDK to run the analysis. The module runs on Android 8, 9, 10, 11, 12 and we expect it to run on Android 13. 
    \item The tool uses \textit{adb} from Android SDK to connect to the virtual device. The application (to be analysed) is installed using adb and inputs to the emulator are provided using adb. 
    \item The virtual device should be running Frida-server to capture API calls called during the execution of the application. 
    \item Static analysis report is required if API calls analysis is performed. DroidDissector requires APKTool if a report is not present to perform static analysis on the application.
\end{itemize}

\subsection{Feature Extraction}

The dynamic analysis module extracts all the major features used in Android malware detection. These include system calls, API calls, logcat files and network traffic. The module allows the user to set the number of virtual events to input to the application and the maximum time the application should run for. We use the monkey tool from Android to input virtual events to imitate normal usage of an application.

\subsubsection*{System Calls}

The tool extracts system calls made by the application using \textit{strace}, a Linux tool to monitor interaction between applications and the Linux kernel. It is essential to collect all the system calls made by the application to have a complete and thorough system calls analysis. We achieve this by running strace on the zygote process. This allows the tool to monitor every process running in the OS, as the Android OS uses the zygote process to start any other process. The tool filters out the system calls made by the application it is monitoring at the end of the analysis. This is different to the approach seen in past works, where researchers run the application first, and then start the system call analysis on the application, which may lead to an incomplete system calls analysis, i.e., some system calls made by the application between starting an application and getting its process ID may be lost. 

Finally, the tool saves the output strace for every application in a separate file.

\subsubsection*{Network Traffic}

Android emulator provides the user with ``tcpdump" option. tcpdump is a network analyzer utility for Linux. DroidDissector uses tcpdump to analyze the network traffic of the application and stores the network traffic in ``pcap" files. The pcap files can then be run on Wireshark \cite{Wireshark} to extract and analyze network traffic of the application. 

DroidDissector saves a separate network traffic for every application. 

\subsubsection*{Android Logcat}

Logcat is a command line tool provided by Android SDK that outputs the logs, system messages on errors and messages written by the programmer during the execution of the application. DroidDissector runs logcat using the adb utility and saves the log files of each application separately for further analysis.  

\subsubsection*{API Calls}

The API calls module is different to every other dynamic analysis module. Android has over 130,000 API calls, making it computationally not feasible to hook each API call. Therefore, the API calls module requires an input from the user; this could either be the static analysis report of the application stating the API calls actually used by the application in the source code, or a pre-defined list of API calls to be tracked by DroidDissector.

Frida \cite{frida} is a tool that allows hooking of API calls and monitors their execution during the run-time of an application. The tool does not miss any API call during the run-time of an application by following the process outlined. The output also includes information on the parameters passed when the application called the API.  Finally, DroidDissector saves all the APIs called during execution in a separate file for further analysis.

\section{Conclusions}

The main aim of DroidDissector is to provide researchers with a powerful, centralized, and integrated tool for extracting static and dynamic features. This is especially important in static analysis, where there are very few, if any, tools available to extract multiple kinds of features from one tool. On the other hand, up to date dynamic analysis tools are not available. DroidDissector runs on the latest version of Android and is easy to set up and use for feature extraction. The list of features extracted by the tool were informed by our extensive literature review. Our aim is to keep improving this tool, adding functionalities to extract more features and integrate more tools. 

\bibliography{bib}
\bibliographystyle{IEEEtran}
\end{document}